\documentclass[%
reprint,
superscriptaddress,
amsmath, amssymb,
aps,
prd,
]{revtex4-2}

\usepackage{graphicx}
\graphicspath{{./figures/}}
\usepackage{xcolor}
\usepackage{ulem}
\usepackage{dcolumn}
\usepackage{bm}
\usepackage{hyperref}
\hypersetup{colorlinks, linkcolor = [rgb]{0, 0, 0.5}, citecolor = [rgb]{0,0.0,0.5}, urlcolor = [rgb]{0,0.0,0.5}}
\usepackage[caption=false]{subfig}
\usepackage{siunitx}
\usepackage[capitalise]{cleveref}
\usepackage{subcaption} 
\usepackage{comment} 
\usepackage{amssymb}
\usepackage{tabularx}
\usepackage{booktabs} 
\newcolumntype{C}{>{\centering\arraybackslash}X} 
\allowdisplaybreaks

\begin{document}

\title{A global analysis of $\pi^0$, $K_S^0$ and $\eta$ fragmentation functions with BESIII data}

\author{Mengyang Li}
\affiliation{Key Laboratory of Atomic and Subatomic Structure and Quantum Control (MOE), Guangdong Basic Research Center of Excellence for Structure and Fundamental Interactions of Matter, Institute of Quantum Matter, South China Normal University, Guangzhou 510006, China}
\affiliation{Guangdong-Hong Kong Joint Laboratory of Quantum Matter, Guangdong Provincial Key Laboratory of Nuclear Science, Southern Nuclear Science Computing Center, South China Normal University, Guangzhou 510006, China}

\author{Daniele Paolo Anderle}
\affiliation{Key Laboratory of Atomic and Subatomic Structure and Quantum Control (MOE), Guangdong Basic Research Center of Excellence for Structure and Fundamental Interactions of Matter, Institute of Quantum Matter, South China Normal University, Guangzhou 510006, China}
\affiliation{Guangdong-Hong Kong Joint Laboratory of Quantum Matter, Guangdong Provincial Key Laboratory of Nuclear Science, Southern Nuclear Science Computing Center, South China Normal University, Guangzhou 510006, China}

\author{Hongxi Xing}
\email{hxing@m.scnu.edu.cn}
\affiliation{Key Laboratory of Atomic and Subatomic Structure and Quantum Control (MOE), Guangdong Basic Research Center of Excellence for Structure and Fundamental Interactions of Matter, Institute of Quantum Matter, South China Normal University, Guangzhou 510006, China}
\affiliation{Guangdong-Hong Kong Joint Laboratory of Quantum Matter, Guangdong Provincial Key Laboratory of Nuclear Science, Southern Nuclear Science Computing Center, South China Normal University, Guangzhou 510006, China}
\affiliation{Southern Center for Nuclear-Science Theory (SCNT), Institute of Modern Physics, Chinese Academy of Sciences, Huizhou 516000, China}

\author{Yuxiang Zhao}
\email{yxzhao@impcas.ac.cn}
\affiliation{Institute of Modern Physics, Chinese Academy of Sciences, Lanzhou 730000, China}
\affiliation{Southern Center for Nuclear-Science Theory (SCNT), Institute of Modern Physics, Chinese Academy of Sciences, Huizhou 516000, China}
\affiliation{University of Chinese Academy of Sciences, Beijing 100049, China}
\affiliation{Key Laboratory of Quark and Lepton Physics (MOE) and Institute of Particle Physics, Central China Normal University, Wuhan 430079, China}

\date{\today}

\date{\today}

\begin{abstract}
In this research, we conduct a global QCD analysis of fragmentation functions (FFs) for neutral pions ($\pi^0$), neutral kaons ($K_S^0$), and eta mesons ($\eta$), utilizing world data of single inclusive hadron production in $e^+e^-$ annihilation involving the most recent BESIII data with low collision energy, to test the operational region of QCD collinear factorization for single inclusive hadron production. We found that the QCD-based analysis at next-to-next-to leading order in perturbative QCD with parameterized higher-twist effects can explain both existing high-energy world data and the BESIII new measurements, while the latter cannot be explained with existing FFs extracted with high-energy data.
To investigate the higher-twist contributions to this discrepancy, a direct functional approach is employed, providing testing framework for characterizing the experimental results over a wide range of energy scales, from low to high, thus extending the classical theoretical models to the BESIII domain.

\end{abstract}

\maketitle
\section{Introduction}
To describe the process of hadronization, fragmentation functions(FFs) \cite{Field:1977fa} are introduced into the framework of Quantum Chromodynamics (QCD). Fragmentation functions characterize the non-perturbative process of a parton, from hard scattering events, evolving into an observed hadron in final state \cite{Collins:1981uk,Collins:1981uw}, where FFs represent the probability density of a parton fragmenting into a specific hadron. Similar to parton distribution functions (PDFs) \cite{Forte:2010dt,Forte:2013wc,Rojo:2015acz}, these functions have been proven to be universal and cannot be computed perturbatively, and can only be extracted from world data through global QCD analysis. These analyses encompass hadron production in electron–positron Single-Inclusive Annihilation (SIA), lepton–nucleon Semi-Inclusive Deep-Inelastic Scattering (SIDIS), and proton–proton (pp) collisions, with the stipulation that the factorization scale must significantly exceed the QCD scale parameter ($\Lambda_{\rm QCD}$). Owing to this rationale and the scarcity of data at low energy scales, QCD global analyses of FFs usually incorporate experiments at high energy scales. Therefore, the acquisition of new data for $\pi^0$, $K_S^0$, and $\eta$ at relatively low scales from BESIII \cite{BESIII:2022zit,BESIII:2024hcs} mandates a critical evaluation of the QCD factorization framework across these energy regions.

Numerous collaborations regularly update and refine their sets of FFs for both light and heavy hadrons, complete with quantified uncertainties. For neutral pions ($\pi^0$), the FFs of their charged counterparts are frequently used as proxies for estimations. Over the recent years, FFs for charged pions at next-to-leading order (NLO) accuracy have been established by several groups, such as DEHSS \cite{deFlorian:2014xna}, HKKS \cite{Hirai:2016loo}, JAM \cite{Sato:2016wqj} and \cite{Gao:2024nkz}. Theoretical investigations into the effects of next-to-next-to-leading order (NNLO) QCD corrections have been conducted by ARS \cite{Anderle:2015lqa}, AKRS \cite{Anderle:2016czy}, NNFF \cite{Bertone:2017tyb}, BSDSV \cite{Borsa:2022vvp} and MAPFF \cite{AbdulKhalek:2022laj}. However, while AKRS incorporated small-$z$ resummation, NNFF \cite{Bertone:2017tyb} accounted for hadron mass corrections, with each group employing distinct initial evolution scales and kinematic constraints on the data. BSDSV \cite{Borsa:2022vvp} and MAPFF \cite{AbdulKhalek:2022laj} NNLO analyses include lepton-nucleon semi-inclusive deep inelastic scattering data. In the case of neutral kaons ($K_S^0$), the literature features contributions from several notable collaborations, including BKK96 \cite{Binnewies:1995kg}, BS \cite{Bourrely:2003wi}, AKK05 \cite{Albino:2005mv}, AKK08 \cite{Albino:2008fy}, and SAK20 \cite{Soleymaninia:2020ahn}, each offering their unique sets of FFs. BKK96 \cite{Binnewies:1995kg} and AKK05 \cite{Albino:2005mv} carried out a comprehensive QCD analysis to derive NLO accuracy FFs for $K_S^0$, utilizing data from electron-positron collisions. AKK08 \cite{Albino:2008fy} revisited FFs for $K_S^0$ considering the effect from hadron mass corrections, including inclusive hadron production measurements from proton-proton collisions at PHENIX, STAR, BRAHMS, and CDF. SAK20 \cite{Soleymaninia:2020ahn} FFs determinations are performed at NNLO accuracy, which is the most recent QCD analysis for the fragmentation functions of $K_S^0$. When it comes to the FFs for the $\eta$ meson, the AESSS \cite{Aidala:2010bn} FFs at NLO accuracy stand as the primary and sole resource. 

Despite significant advancements, our current comprehension of the FFs for $\pi^0$, $K_S^0$, and $\eta$ suffers from describing low scale data points. The predictions from existing FFs for $\pi^0$, $K_S^0$, and $\eta$ significantly deviate from experimental results within the low-energy range of the BESIII experiment \cite{BESIII:2022zit,BESIII:2024hcs}, one significant potential reason for this discrepancy may be that the FF analyses predominantly utilize SIA data with $\sqrt{s}$ values above $9$ GeV, a region where higher twist effects, suppressed by $Q^2$, are often neglected.

In this study, we make a comprehensive global QCD analysis of single inclusive production for $\pi^0$, $K_S^0$, and $\eta$, with existing SIA world data and the latest low $Q$ scale BESIII data \cite{BESIII:2022zit,BESIII:2024hcs}, to test the leading twist QCD collinear factorization working region. In addition, a parameterized functional approach is adopted to explore the higher twist contribution. This approach offers a testing framework for describing experimental outcomes across a broad spectrum of energy levels, from low to high, extending the classical theoretical descriptions to BESIII region.

The remainder of the paper is organized as follows. We first introduce the QCD factorization framework for SIA in Sec.~\ref{sec-SIA}, we then present the fitting framework in our analysis in Sec.~\ref{sec-fit-overall} followed by the final result in Sec.~\ref{sec-res}. Finally, a summary is given in Sec.\ref{sec-summary}.

\section{Physical observable}
\label{sec-SIA}
In order to test the operational region of leading twist QCD collinear factorization for hadron production, and study the influence of higher twist effect at low collisional energies, we have chosen to analyze the SIA process, which is preferred 
because its interpretation does not require the simultaneous knowledge of PDFs, making it the theoretically cleanest process for FF studies. The cross section for the single-inclusive hadron production in electron-positron annihilation ($e^++e^- \to h+X$), normalized to the total cross section $\sigma_{tot}$ ($e^++e^- \to X$), can be schematically written as 
\begin{equation}
\begin{aligned}
    \frac{1}{\sigma_{tot}}\frac{d\sigma^h}{dz} = \frac{1}{\sigma_{tot}}\sum_{k=T,L}\frac{d\sigma^h_k}{dz},
    \label{eq-1}
\end{aligned}
\end{equation}
where $z=2P_h\cdot q/Q^2$ is the scaling variable ($z=2E_h/\sqrt{s}$ in electron-positron center of mass frame), $P_h$ and $q$ are the four-momenta of the observed hadron and the time-like $\gamma/Z$ boson, respectively, and $Q^2=q^2=s$ with $\sqrt{s}$ being the center of mass energy. $T,~L$ denote the transverse and longitudinal polarization of the exchanged $\gamma/Z$ boson.

According to QCD factorization theorem for SIA process when $Q\gg \Lambda_{\rm QCD}$, the transverse (T) and longitudinal (L) part in Eq.(\ref{eq-1}) can be written as a convolution of perturbative coefficient functions $C_i^k$ and non-perturbative fragmentation functions (FFs) $D_i^h$,
\begin{equation}
\begin{aligned}
     \frac{1}{\sigma_{tot}}\frac{d\sigma^h_k}{dz}=\frac{1}{\sigma_{tot}}\sum_i C_i^k\left(z,Q^2,\mu^2\right) \otimes D_i^h(z,\mu^2).
    \label{eq-lt}
\end{aligned}
\end{equation}
In our calculation, both factorization and renormalization scales are set equal to the center of mass energy of the collision, $\mu_R=\mu_F=\sqrt{s}=Q$. The symbol $\otimes$ denotes the standard convolution integral defined as 
\begin{equation}
\begin{aligned}
    [f\otimes g](z)=\int_0^1dx\int_0^1dy f(x)g(y)\delta(z-xy).
\end{aligned}
\end{equation}
In pQCD, the coefficient functions can be calculated as a perturbative series in $a_s=\alpha_s/4\pi$ with $\alpha_s$ the QCD running coupling,
\begin{equation}
\begin{aligned}
    C_i^k=C_i^{k,(0)}+a_s C_i^{k,(1)}+a_s^2C_i^{k,(2)}+\dots .
\end{aligned}
\end{equation}
In order to minimize the uncertainties from higher order QCD corrections, we use the highest precision results for all the perturbative coefficients up to date, i.e., NNLO at $\mathcal{O}(a_s^2)$, where the expressions for $\sigma_{tot}$ and the coefficient functions can be found in Refs. \cite{Chetyrkin:1979bj,Rijken:1996vr, Rijken:1996ns, Mitov:2006wy, Blumlein:2006rr}.

In our analysis, fragmentation functions for $\pi^0$, $K_S^0$, and $\eta$ are derived from SIA, integrating both the comprehensive global datasets and the latest BESIII data \cite{BESIII:2022zit,BESIII:2024hcs}. The global dataset encompasses a wide range of center-of-mass energies, from 9.46 GeV at ARGUS, to 91.2 GeV at facilities such as ALEPH and OPAL. In addition to inclusive measurements, our study integrates flavor-tagged data from the TPC, DELPHI, and SLD experiments, facilitating the differentiation between contributions from different quark and gluon types. Additionally, the existing predictions for $\pi^0$ are derived from $\pi^{\pm}$ fragmentation functions. To maintain consistency with past research and enhance the dataset, our analysis includes world data for both neutral and charged pions in SIA.

The latest BESIII data \cite{BESIII:2022zit,BESIII:2024hcs} expands the energy scale down to 2.0 GeV, a regime that challenges leading twist QCD factorization for single inclusive hadron production, thereby enabling an exploration of higher twist effects to augment the understanding of QCD factorization.

The datasets employed in this study are summarized in  Sec.~\ref{sec-res}.  For each dataset, we specify the experiment's name, related references, and the number of data points included in the fit. In the small $z$ region, soft gluon effects cause the DGLAP evolution equations to become unstable, leading to discrepancies between the theoretical models and experimental data. Consequently, all theoretical models limit their analysis to data points where $z \geq z_{min}$, with $z_{min}$ representing the lower $z$-bound. In our analysis, we constrain the data to $z \geq z_{min} = 0.05$ and apply an upper cut at $z < 0.95$ to mitigate the significant enhancement due to threshold logarithms $\propto \log(1 - z)$.

\section{OUTLINE OF THE ANALYSIS}
\label{sec-fit-overall}

\subsection{Fitting framework}
\label{sec-fit}

The fitting framework we employ is grounded in the methodology outlined in Ref. \cite{Anderle:2017cgl}, where the fitting process is performed in Mellin space. Additionally, the evolution of FFs across varying energy scales, denoted by $Q$, is regulated by the Dokshitzer-Gribov-Lipatov-Altarelli-Parisi (DGLAP) equations \cite{Gribov:1972ri,Lipatov:1974qm,Altarelli:1977zs,Dokshitzer:1977sg}:
\begin{equation}
\begin{aligned}
    \frac{\partial}{\partial {\rm ln} \mu^2}D_i^h(z,\mu)=\sum_j [P_{ji} \otimes D_j^h] (z,\mu),
\end{aligned}
\end{equation}
where $P_{ji}$ are the time-like splitting functions, and $z$ is the fraction of the four-momentum of the parton taken by the identified hadron.

Our primary objective is to investigate the FFs of $\pi^0$, $K_S^0$, and $\eta$, including the exploration of higher twist effects. We parameterize the FFs at an initial scale $\mu_0=1$ GeV, which is below the lowest energy region encountered in BESIII collisions. We take the parameterization form of FFs at initial scale $\mu_0$ following the approach in the series of DSS global QCD analyses \cite{Anderle:2015lqa,deFlorian:2007aj,deFlorian:2017lwf,deFlorian:2007ekg}:
\begin{equation}
\begin{aligned}
    D_i^{h}(z,\mu_0)=\frac{N_iz^{\alpha_i}(1-z)^{\beta_i}[1+\gamma_i(1-z)^{\delta_i}]}{B[2+\alpha_i,\beta_i+1]+\gamma_iB[2+\alpha_i,\beta_i+\delta_i+1]}.
    \label{D-ini}
\end{aligned}
\end{equation}
$B[a,b]$ denotes the Euler Beta function, with $a$ and $b$ chosen such that $N_i$ is normalized to the second moment $\int_0^1 zD_i^h(z,\mu_0)dz$ of the FFs.
Given the limitation of using SIA data, which precludes differentiation between quark and antiquark fragmentation functions, our analysis adopts the quark combination $q^+=q+\bar{q}$ for the parametrization. In Equation (\ref{D-ini}), the index $i$ represents the quark combinations $u^+$, $d^+$, $s^+$, $c^+$, $b^+$, and the gluon $g$.

\begin{table}[h]
\centering
\caption{Parameters describing the NNLO FFs ($i^+=i+\bar{i}$) for $\pi^0$. Inputs for light and gluon FFs are set at the initial scale $\mu_0 = 1.0$ GeV. Inputs for the charm and bottom FFs refer to $\mu=m_c$ and $\mu=m_b$, respectively.}
\label{tab:par_1}
\begin{tabularx}{0.45\textwidth}{@{}lCCCCC@{}}
\toprule
Parameter & N & $\alpha$ & $\beta$ & $\gamma$ & $\delta$ \\
\midrule
$u^+=d^+$ & $N_{u^+}$& $\alpha_{u^+}$ & $\beta_{u^+}$ & $\gamma_{u^+}$ & $\delta_{u^+}$ \\
$s^+$ & $N_{s^+}$& $\alpha_{s^+}$ & $\beta_{s^+}$ & $\gamma_{s^+}$ & $\delta_{s^+}$ \\
$g$   & $N_{g}$  & $\alpha_{g}$   & $\beta_{g}$   & $\gamma_{g}$   & $\delta_{g}$   \\
$c^+$ & $N_{c^+}$& $\alpha_{c^+}$ & $\beta_{c^+}$ & $\gamma_{c^+}$ & $\delta_{c^+}$ \\
$b^+$ & $N_{b^+}$& $\alpha_{b^+}$ & $\beta_{b^+}$ & $\gamma_{b^+}$ & $\delta_{b^+}$ \\
\bottomrule
\end{tabularx}
\end{table}

\begin{table}[h]
\centering
\caption{Same as Tab.\ref{tab:par_1}, but for $K_S^0$.}
\label{tab:par_2}
\begin{tabularx}{0.45\textwidth}{@{}lCCCCC@{}}
\toprule
Parameter & N & $\alpha$ & $\beta$ & $\gamma$ & $\delta$ \\
\midrule
$u^+$ & $N_{u^+}$& $\alpha_{u^+}$ & $\beta_{u^+}$ & - & - \\
$d^+$ & $N_{d^+}$& $\alpha_{d^+}$ & $\beta_{d^+}$ & - & - \\
$s^+$ & $N_{s^+}$& $\alpha_{s^+}$ & $\beta_{s^+}$ & - & - \\
$g$   & $N_{g}$  & $\alpha_{g}$   & $\beta_{g}$   & - & - \\
$c^+$ & $N_{c^+}$& $\alpha_{c^+}$ & $\beta_{c^+}$ & - & - \\
$b^+$ & $N_{b^+}$& $\alpha_{b^+}$ & $\beta_{b^+}$ & - & - \\
\bottomrule
\end{tabularx}
\end{table}

\begin{table}[h]
\centering
\caption{Same as Tab.\ref{tab:par_1}, but for $\eta$.}
\label{tab:par_3}
\begin{tabularx}{0.45\textwidth}{@{}l*{5}{C}@{}}
\toprule
Parameter & N & $\alpha$ & $\beta$ & $\gamma$ & $\delta$ \\
\midrule
$u^+=d^+=s^+$ & $N_{u^+}$& $\alpha_{u^+}$ & $\beta_{u^+}$ & $\gamma_{u^+}$ & $\delta_{u^+}$ \\
$g$ & $N_{g}$  & $\alpha_{g}$   & $\beta_{g}$ & - & - \\
$c^+$ & $N_{c^+}$& $\alpha_{c^+}$ & $\beta_{c^+}$ & - & - \\
$b^+$ & $N_{b^+}$& $\alpha_{b^+}$ & $\beta_{b^+}$ & - & - \\
\bottomrule
\end{tabularx}
\end{table}
We now outline the assumptions underlying our parametrizations. Based on the isospin symmetry of different mesons, various strategies are employed for the parametrization of fragmentation functions for $\pi^0$, $K_S^0$, and $\eta$. In the case of the neutral pion ($\pi^0$), we posit identical fragmentation functions for up and down quarks, denoted as $D^{\pi^0}_{u^+} = D^{\pi^0}_{d^+}$. For the $K_S^0$ meson, we adhere to the fragmentation function scheme outlined in AKK05 \cite{Albino:2005mv}, and followed in AKK08 \cite{Albino:2008fy} and SAK20 \cite{Soleymaninia:2020ahn}, where $u^+,d^+,s^+$ are parametrized independently. Considering the number of experimental data points from $e^+e^-$ annihilation to determine the $K_S^0$ FFs is rather limited, we fix $\gamma_i, \delta_i$ to zero. Regarding the $\eta$ meson, we adopt the AESSS \cite{Aidala:2010bn} scheme, which posits that all light quark flavors have equivalent fragmentation functions, namely $D_{u^+}^\eta = D_{d^+}^\eta = D_{s^+}^\eta$. For the same reasons as those governing $K_S^0$, we impose constraints on the parameters $\gamma_{c^+,b^+,g}$ and $\delta_{c^+,b^+,g}$ for $\eta$. In the scale evolution of the FFs, charm and bottom quarks are included in the scale evolution above $\mu = m_{c}$ and $\mu = m_{b}$, where $m_c = 1.43$ GeV and $m_b = 4.3$ GeV represent the masses of the charm and bottom quarks, respectively. After all these considerations, the free parameters for $\pi^0$, $K_S^0$, and $\eta$ are summarized in Tables \ref{tab:par_1}, \ref{tab:par_2}, and \ref{tab:par_3}.

\begin{figure*}[t] 
    \centering 
    \includegraphics[width=0.85\textwidth]{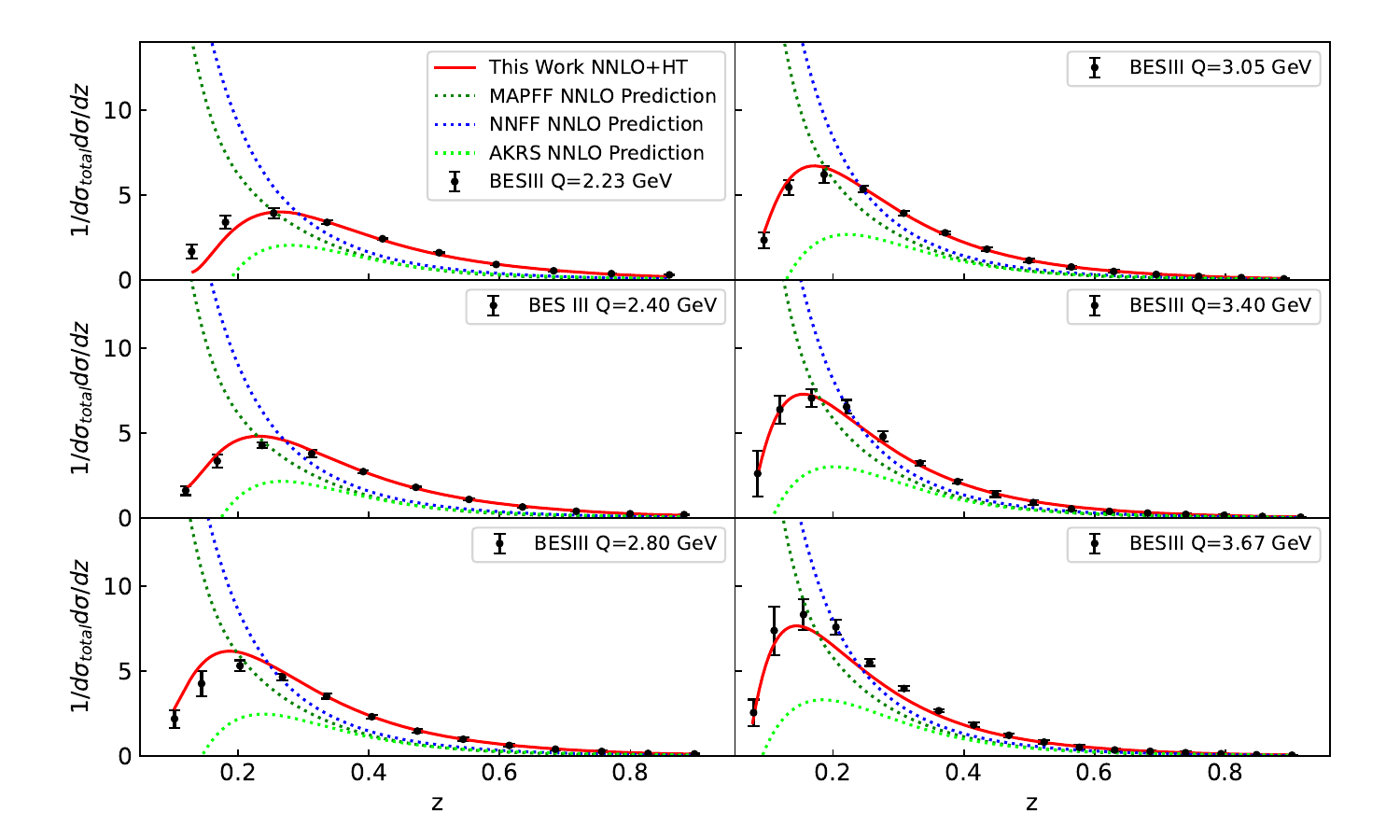} 
    \caption{Analysis of $\pi^0$ BESIII \cite{BESIII:2022zit} datasets alongside our best-fit results, compared with the NNLO predictions from NNFF \cite{Bertone:2017tyb}, MAPFF \cite{AbdulKhalek:2022laj}, and AKRS \cite{Anderle:2016czy}.} 
    \label{fig:pi0_history} 
\end{figure*}

\begin{figure*}[t] 
    \centering 
    \includegraphics[width=0.85\textwidth]{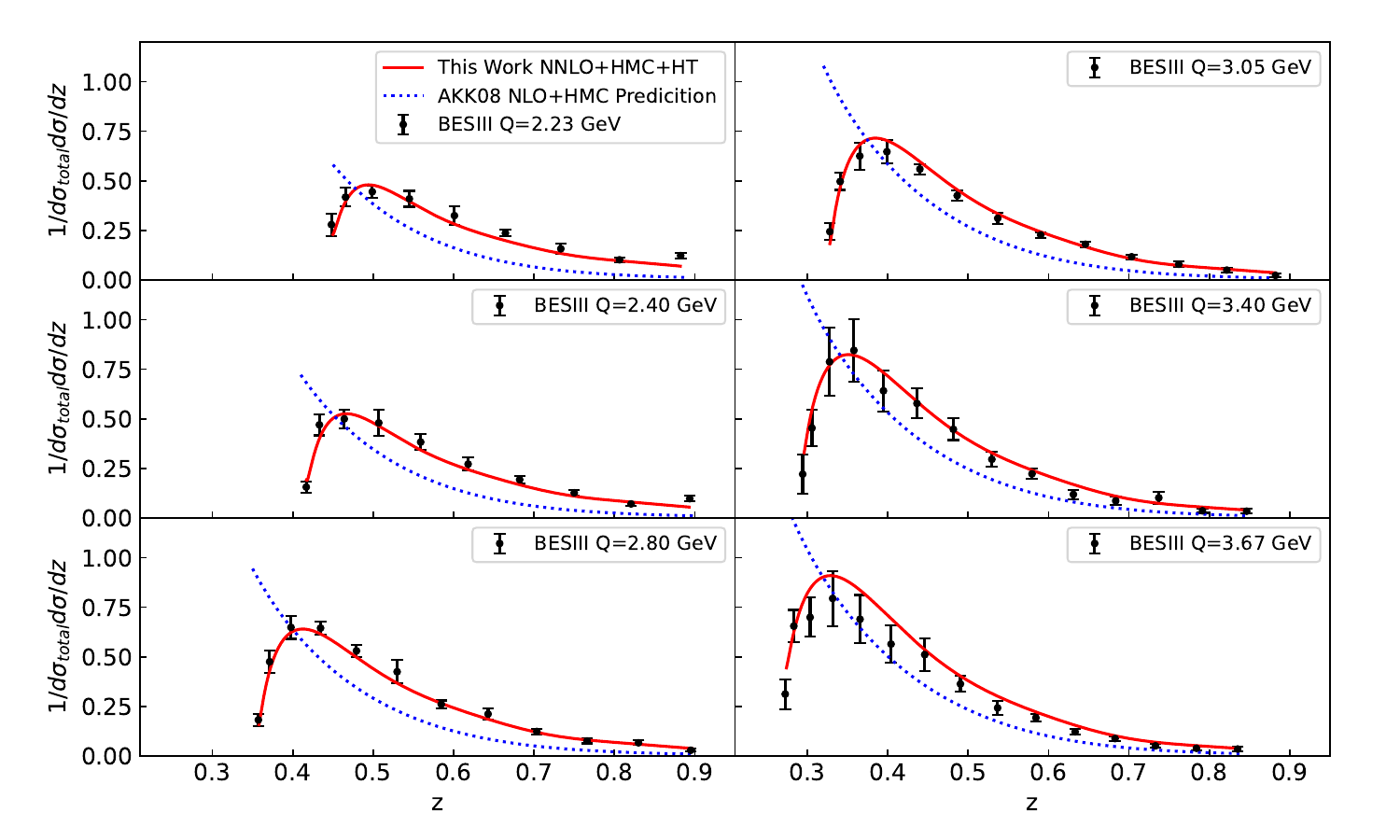} 
    \caption{Analysis of $K_S^0$ BESIII \cite{BESIII:2022zit} datasets alongside our best-fit results, compared with the NLO predictions from AKK08 \cite{Albino:2008fy}.} 
    \label{fig:ks_history} 
\end{figure*}

\begin{figure*}[t] 
    \centering 
    \includegraphics[width=0.85\textwidth]{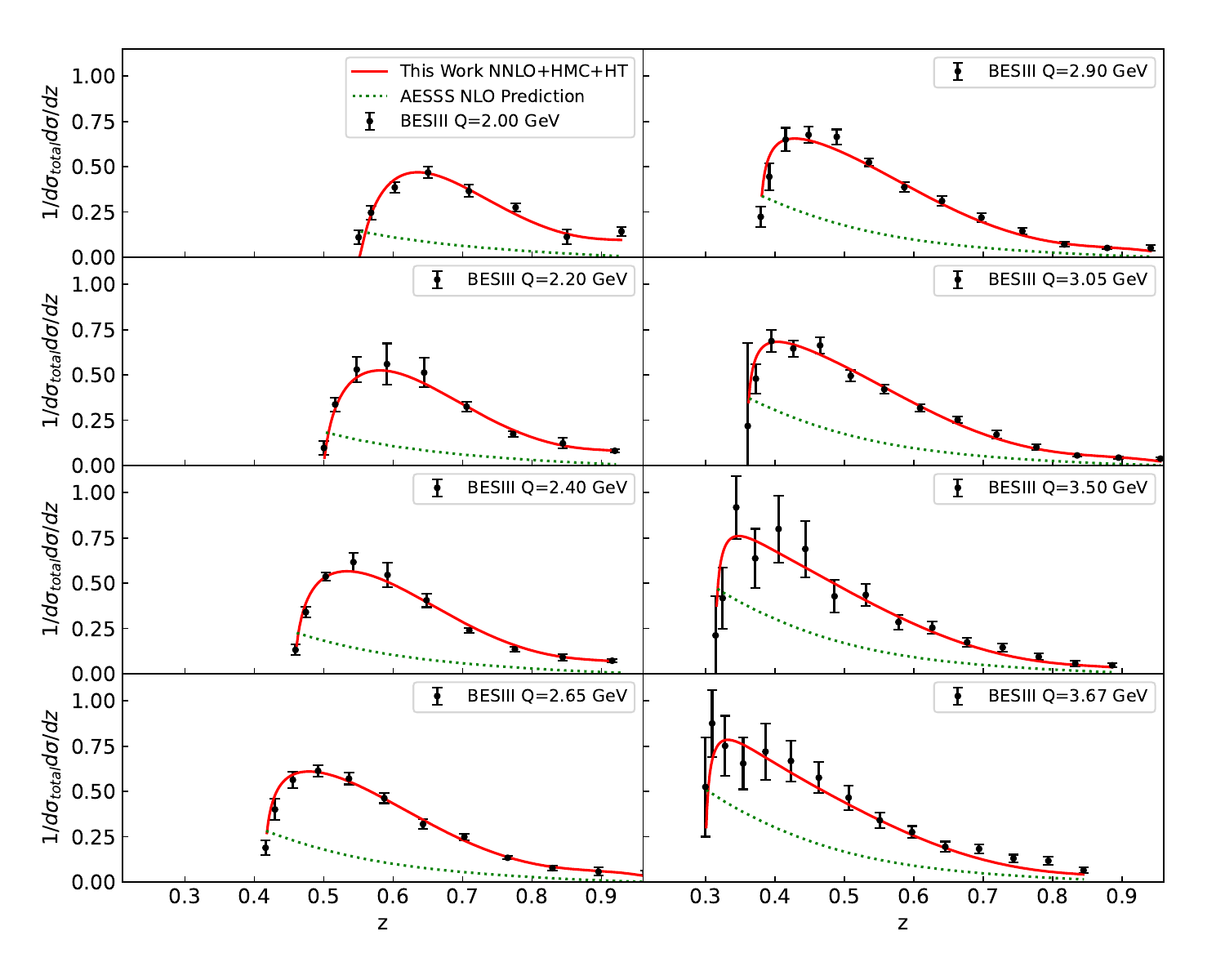} 
    \caption{Analysis of $\eta$ BESIII \cite{BESIII:2024hcs} datasets alongside our best-fit results, compared with the NLO predictions from AESSS \cite{Aidala:2010bn}.} 
    \label{fig:eta_history} 
\end{figure*}

Parameters characterizing the FFs of quarks and gluons into neutral mesons, as introduced in Eq. (\ref{D-ini}), are determined through standard $\chi^2$ minimization using world data \cite{Belle:2013lfg,BaBar:2013yrg,TPCTwoGamma:1988yjh,SLD:1998coh,ALEPH:1994cbg,OPAL:1994zan,DELPHI:1998cgx,ARGUS:1989orf,TASSO:1981rff,TASSO:1986tou,CELLO:1983mif,CELLO:1989byk,TPCTwoGamma:1984eyl,JADE:1985bzp,JADE:1989ewf,ALEPH:1996oqp,ALEPH:1999udi,OPAL:1998enc,L3:1991nwl,L3:1994gkb,TASSO:1984nda,TASSO:1989jyt,Derrick:1985wd,TPCTwoGamma:1984eoj,Schellman:1984yz,CELLO:1989adw,TOPAZ:1994voc,DELPHI:1994qgk,OPAL:1999zfe,DELPHI:2000ahn,ARGUS:1989orf,HRS:1987aky,Wormser:1988ru,JADE:1985bzp,JADE:1989ewf,CELLO:1989byk,ALEPH:1992zhm,ALEPH:1999udi,ALEPH:2001tfk,L3:1992pbe,L3:1994gkb,OPAL:1998enc} and BESIII data \cite{BESIII:2022zit,BESIII:2024hcs}. The minimized $\chi^2$ value is defined by the equation \begin{equation} 
\chi^2 = \sum^N_{j=1}\frac{(T_j-E_j)^2}{\Delta E_j^2}, 
\label{chi2} 
\end{equation} 
where $E_j$ denotes the experimentally measured value, $\Delta E_j$ is the associated uncertainty, and $T_j$ is the theoretical predictions  for the parameters in Eq. (\ref{D-ini}), at a specified order in $\alpha_s$.
For the experimental uncertainties $\Delta E_j$ we consider the statistical and systematic errors in quadrature for the time being.

Considering the relatively low energy scale at BESIII, one might also need to consider the hadron mass corrections \cite{Albino:2005gd,Accardi:2014qda,MoosaviNejad:2015lgp}. In the presence of hadron mass effect, the scaling variable needs to be modified from $z=2E_h/\sqrt{s}$ to a specific choice of scaling variable $\xi$ defined as a light-cone scaling. It is given by,
\begin{equation}
\begin{aligned}
    \xi = \frac{z}{2}\left(1+\sqrt{1-\frac{4m_h^2}{Q^2 z^2}}\right).
\end{aligned}
\end{equation}
Consequently, the differential cross section in the presence of hadron mass effect for SIA process need to be modified as 
\begin{equation}
\begin{aligned}
    \frac{d\sigma^h}{dz}=\frac{1}{1-\frac{m_h^2}{Q^2\xi^2}}\frac{d\sigma^h}{d\xi}.
\end{aligned}
\end{equation}

One notices that hadron mass corrections increase as either $z$ or $\sqrt{s}$ decreases, or when $m_h$ is increased. Therefore, in the low-scale kinematic region, these corrections for $K_S^0$ and $\eta$ can become significant. For example, hadron-mass corrections remain below $10\%$ for all hadronic species at $z = 0.1$ and $Q=M_Z$, but increase to $30\%$ or higher for $K_S^0$ and $\eta$ at $z = 0.1$ and $Q = 10$ GeV. In our analysis, we incorporate hadron-mass corrections for $K_S^0$ and $\eta$ to enhance the theoretical accuracy for the BESIII \cite{BESIII:2022zit,BESIII:2024hcs} data description.

\subsection{Higher-twist effect evaluation}
Analysis of BESIII data \cite{BESIII:2022zit,BESIII:2024hcs}, as presented in Fig.\ref{fig:pi0_history}, \ref{fig:ks_history}, and \ref{fig:eta_history}, utilizing predictions from NNFF \cite{Bertone:2017tyb}, MAPFF \cite{AbdulKhalek:2022laj}, AKRS \cite{Anderle:2016czy}, AKK08 \cite{Albino:2008fy}, and AESSS \cite{Aidala:2010bn}, reveals significant tension. Traditional methods of incorporating BESIII \cite{BESIII:2022zit,BESIII:2024hcs} experiments into the datasets have been less effective in obtaining a robust solution, invariably resulting in a large $\chi^2$, thereby prompting the need for the inclusion of higher-twist effects in the BESIII region. 

Considering the lack of higher-twist contributions in SIA calculations, our analysis follows the methodology of Accardi et al. \cite{Accardi:2009br}, enhancing it to incorporate higher twist effects by parameterizing the corrections with a phenomenological $z$-dependent function:
\begin{equation}
\begin{aligned}
    \frac{d\sigma^h_k}{dz}=\frac{d\sigma^{h,LT}_k}{dz}\left[1+\frac{C_{T4}(z)}{Q^2}+\frac{C_{T6}(z)}{Q^4}+\dots\right],
\end{aligned}
\end{equation}
where $d\sigma^{h,LT}_k$ denotes the leading twist contribution shown in Eq.(\ref{eq-lt}), the second and third term in the above equation are the corresponding twist-4 and twist-6 corrections to leading twist, which are suppressed in $\frac{1}{Q^2}$ and $\frac{1}{Q^4}$, respectively. The higher twist coefficient functions are parameterized by a polynomial function as 
\begin{equation}
\label{EQ.HT}
\begin{aligned}
    C_{T4}(x)=h_0x^{h_1}(1+h_2x),\\
    C_{T6}(x)=h_3x^{h_4}(1+h_5x),
\end{aligned}
\end{equation}
with $h_0,h_1,h_2,h_3,h_4$ and $h_5$ as free parameters. In this study, we adopt $C_{T4}$ and $C_{T6}$, which effectively describe the data points. More complex parametrizations require extensive data analysis, which are beyond the scope of this paper. Consequently, the exploration of higher parameterizations is reserved for future research, in anticipation that advancements in scientific methods will permit a more comprehensive investigation.

\section{RESULTS}
\label{sec-res}

\begin{figure*}[t] 
    \centering 
    \includegraphics[width=0.8\textwidth]{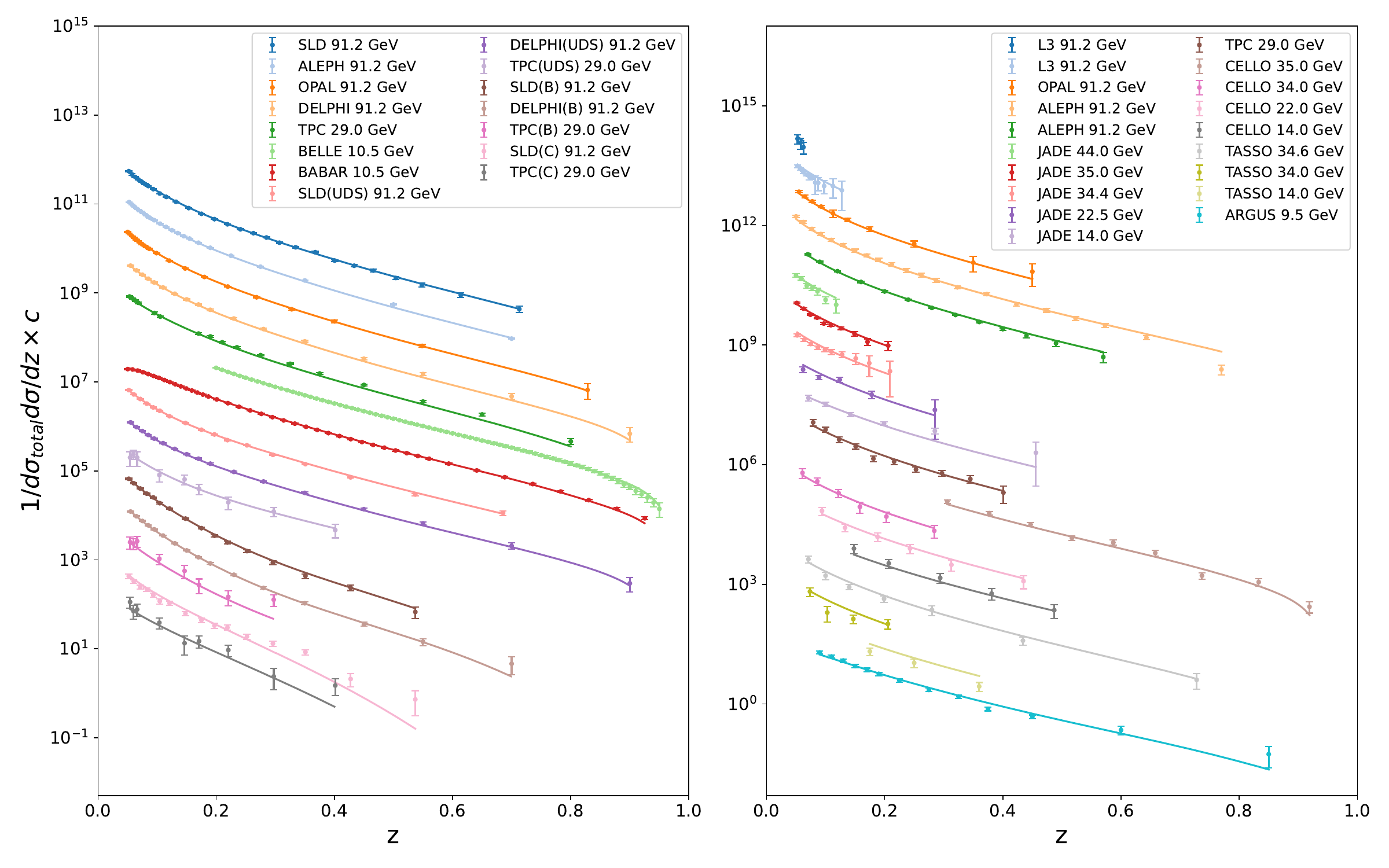} 
    \caption{Comparison of the included world datasets (left for $\pi^{\pm}$, right for $\pi^0$) with the corresponding NNLO theoretical predictions using our best-fit $\pi^0$ NNLO FFs.  The distributions have been scaled by $c=10^i$ with i ranging from TPC(C) to SLD.} 
    \label{fig:my_label1} 
\end{figure*}

\begin{figure*}[] 
    \centering 
    \includegraphics[width=0.7\textwidth]{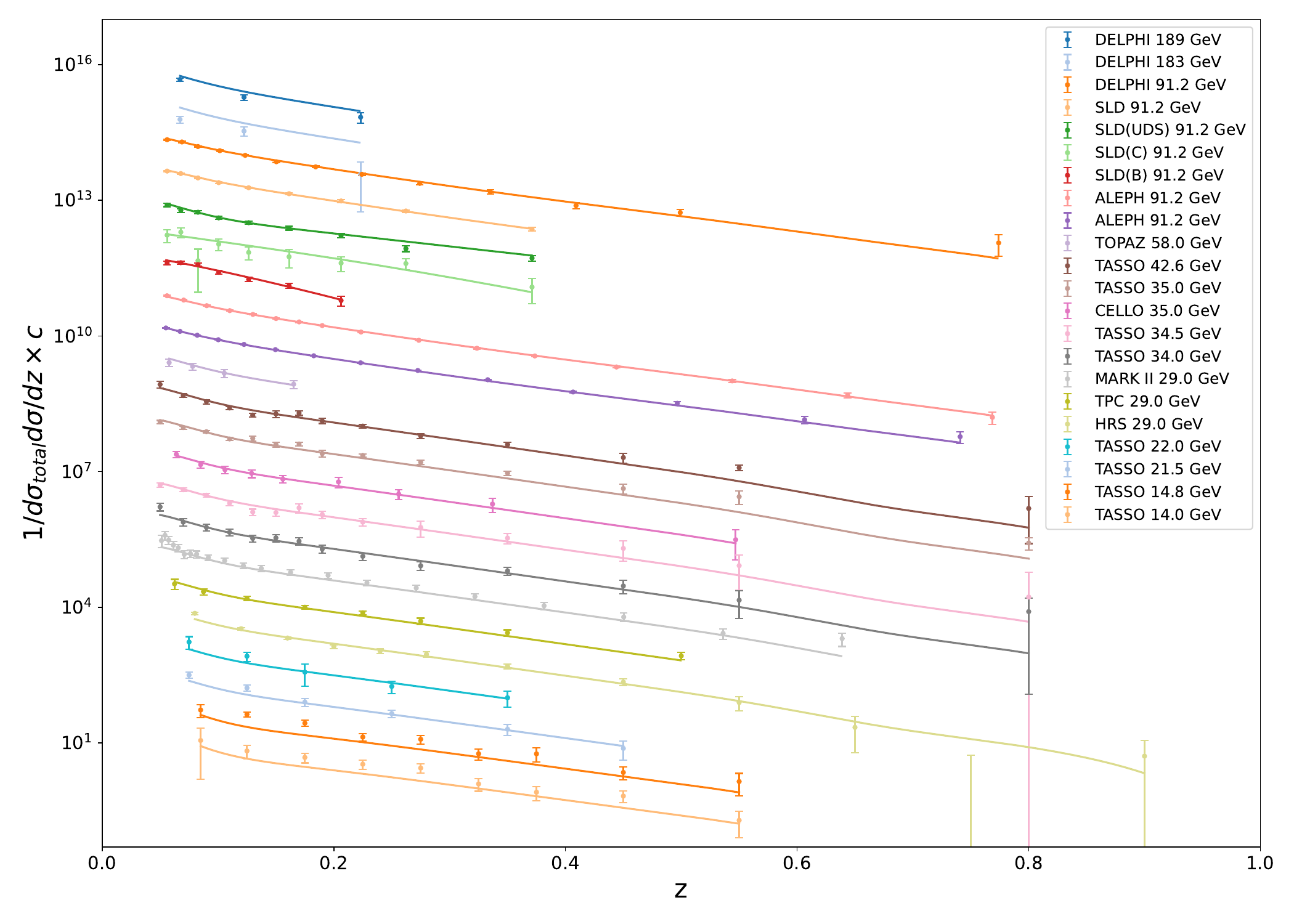} 
    \caption{Same as Fig.\ref{fig:my_label1}, but for $K_S^0$.} 
    \label{fig:my_label2} 
\end{figure*}

\begin{figure*}[] 
    \centering 
    \includegraphics[width=0.7\textwidth]{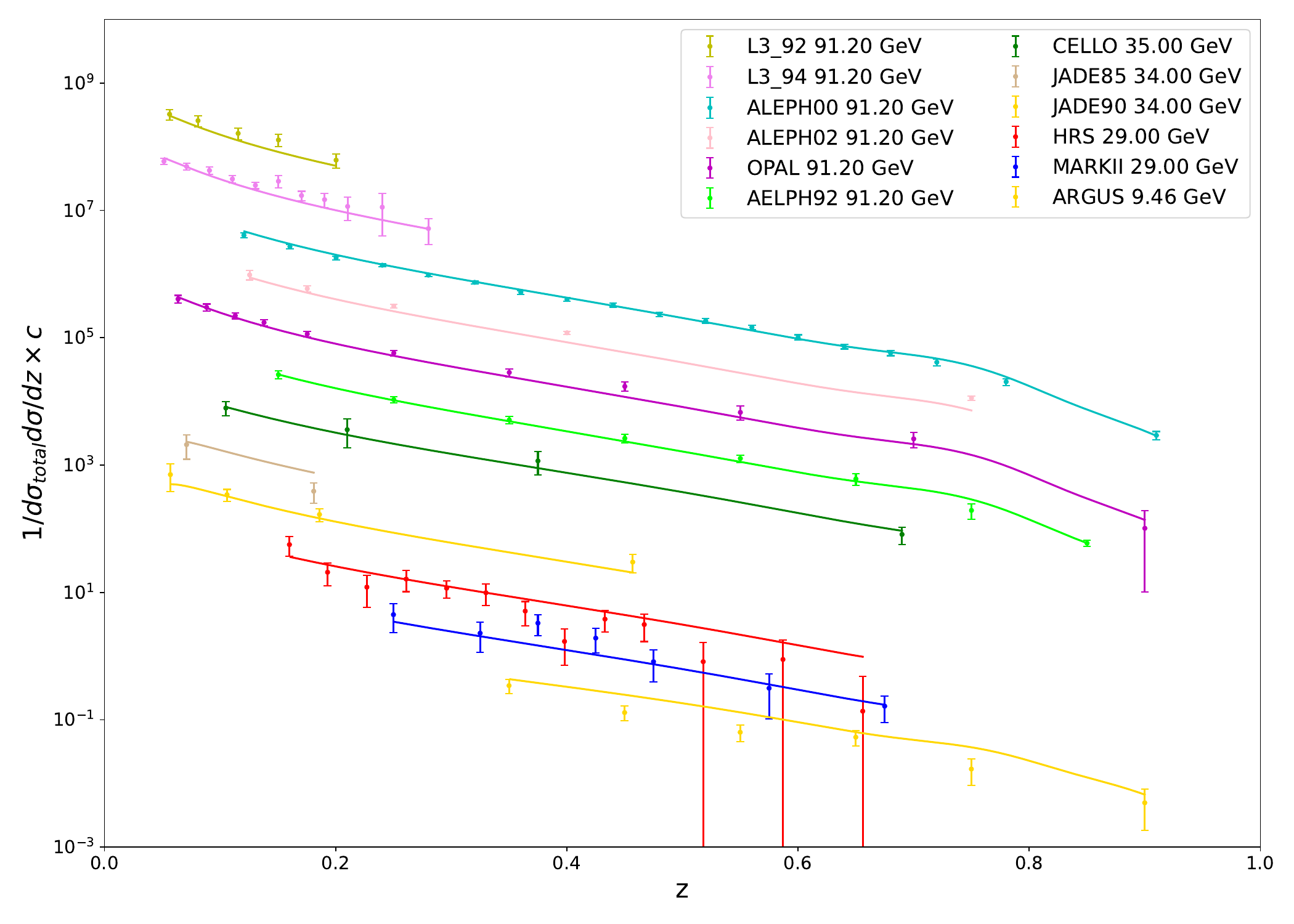} 
    \caption{Same as Fig.\ref{fig:my_label1}, but for $\eta$.} 
    \label{fig:my_label3} 
\end{figure*}

In this section, we present the main results of this work. We first test the global analysis to the SIA data for $\pi^0$, $K^0_S$, and $\eta$ using fixed order pQCD calculation at NNLO without higher twist effect. It turns out that without the BESIII low energy data, we can reproduce the similar $\chi^2$ results in Refs. \cite{Anderle:2016czy, Albino:2008fy, Aidala:2010bn} and a satisfactory fit is obtained for $\pi^0$, $K^0_S$, and $\eta$. However, such a satisfactory convergence of the fit is spoiled when BESIII low energy data is included, even with hadron mass corrections. This observation strongly motivates us to include both mass corrections and higher twist effect, which eventually yield $\chi^2/N_{\text{dp}}$ values of 1.55, 1.47, and 1.52 for $\pi^0$, $K_S^0$, and $\eta$, respectively, indicating a satisfactory fit quality. 

In the following, we discuss the quality of the fits and compare our predictions to the included datasets. The overall statistical quality of our fit, as quantified by the $\chi^2$ per data point ($\chi^2/N_{dp}$), for both individual and combined datasets, is summarized in Tables \ref{tab:my_table1}, \ref{tab:my_table2}, and \ref{tab:my_table3}.

\begin{table}[h]
\centering
\small 
\centering
\caption{The list of input datasets in analyses of $\pi^0$, both $\pi^{\pm}$ and $\pi^0$ data are included. For each dataset, we indicate the corresponding center-of-mass energy $\sqrt{s}$ and the value of $\chi^2$ per data point for the individual dataset in our best-fit. The total values of $\chi^2/N_{dp}$ have been presented as well.}
\begin{tabularx}{0.45\textwidth}{@{}lCCCc@{}} 
\toprule
Exp($\pi^{\pm}$)                          & $\sqrt{s}$ [GeV]    & $N_{dp}$   & $\chi^2/N_{dp}$\\ \midrule
BELLE \cite{Belle:2013lfg} & 10.5 & 76 & 0.23 \\
BABAR \cite{BaBar:2013yrg} & 10.5 & 44 & 1.08 \\
TPC \cite{TPCTwoGamma:1988yjh} & 29.0 & 17 & 4.79 \\
TPC(UDS) \cite{TPCTwoGamma:1988yjh} & 29.0 & 9 & 0.21 \\
TPC(C) \cite{TPCTwoGamma:1988yjh} & 29.0 & 9 & 0.75 \\
TPC(B) \cite{TPCTwoGamma:1988yjh} & 29.0 & 9 & 1.08 \\
SLD \cite{SLD:1998coh}  & 91.2 & 28 & 1.62 \\
ALEPH \cite{ALEPH:1994cbg} & 91.2 & 22 & 3.07 \\
OPAL \cite{OPAL:1994zan} & 91.2 & 21 & 1.07 \\
DELPHI \cite{DELPHI:1998cgx} & 91.2 & 17 & 1.60 \\
DELPHI(UDS) \cite{DELPHI:1998cgx} & 91.2 & 17 & 0.71 \\
DELPHI(B) \cite{DELPHI:1998cgx} & 91.2 & 17 & 0.31 \\
SLD(UDS) \cite{SLD:1998coh} & 91.2 & 17 & 1.25 \\
SLD(C) \cite{SLD:1998coh} & 91.2 & 17 & 2.92 \\
SLD(B) \cite{SLD:1998coh} & 91.2 & 17 & 0.73 \\
\midrule
Exp($\pi^0$)                          & $\sqrt{s}$ [GeV]    & $N_{dp}$   & $\chi^2/N_{dp}$\\ 
\midrule
ARGUS \cite{ARGUS:1989orf} & 9.46 & 13 & 1.90 \\
TASSO \cite{TASSO:1981rff} & 14.0 & 4 & 7.44 \\
TASSO \cite{TASSO:1981rff} & 34.0 & 4 & 2.70 \\
TASSO \cite{TASSO:1986tou} & 34.6 & 7 & 1.51 \\
CELLO \cite{CELLO:1983mif} & 14.0 & 6 & 3.28 \\
CELLO \cite{CELLO:1983mif} & 22.0 & 7 & 0.63 \\
CELLO \cite{CELLO:1983mif} & 34.0 & 7 & 0.49 \\
CELLO \cite{CELLO:1989byk} & 35.0 & 9 & 2.03 \\
TPC \cite{TPCTwoGamma:1984eyl} & 29.0 & 10 & 1.35 \\
JADE \cite{JADE:1985bzp} & 14.0 & 6 & 1.63 \\
JADE \cite{JADE:1985bzp} & 22.5 & 5 & 1.97 \\
JADE \cite{JADE:1985bzp} & 34.4 & 10 & 2.13 \\
JADE \cite{JADE:1989ewf} & 35.0 & 10 & 3.45 \\
JADE \cite{JADE:1989ewf} & 44.0 & 7 & 1.59 \\
ALEPH \cite{ALEPH:1996oqp} & 91.2 & 21 & 2.77 \\
ALEPH \cite{ALEPH:1999udi} & 91.2 & 13 & 1.75 \\
OPAL \cite{OPAL:1998enc}     & 91.2 & 10 & 0.51 \\
L3 \cite{L3:1991nwl} & 91.2 & 3 & 0.87 \\
L3 \cite{L3:1994gkb} & 91.2 & 12 & 2.50 \\
BESIII \cite{BESIII:2022zit} & 2.23 & 10 & 3.33 \\
BESIII \cite{BESIII:2022zit} & 2.40 & 11 & 1.87 \\
BESIII \cite{BESIII:2022zit} & 2.80 & 13 & 1.07 \\
BESIII \cite{BESIII:2022zit} & 3.05 & 14 & 0.55 \\
BESIII \cite{BESIII:2022zit} & 3.40 & 16 & 0.55 \\
BESIII \cite{BESIII:2022zit} & 3.67 & 17 & 2.86 \\
\midrule
TOTAL                             &                &  582 &  1.55\\ 
\bottomrule
\end{tabularx}
\label{tab:my_table1} 
\end{table}

\begin{table}[!h]
\centering
\caption{Same as Tab.\Ref{tab:my_table1}, but for $K_S^0$.}
\begin{tabularx}{0.45\textwidth}{@{}lCCCc@{}} 
\toprule
Exp($K_S^0$)                          & $\sqrt{s}$ [GeV]    & $N_{dp}$   & $\chi^2/N_{dp}$\\
\midrule
TASSO \cite{TASSO:1984nda} & 14.0 & 9 & 1.51 \\
TASSO \cite{TASSO:1989jyt} & 14.8 & 9 & 3.32 \\
TASSO \cite{TASSO:1989jyt} & 21.5 & 6 & 1.14 \\
TASSO \cite{TASSO:1984nda} & 22.0 & 6 & 0.67 \\
HRS \cite{Derrick:1985wd} & 29.0 & 12 & 2.41 \\
TPC \cite{TPCTwoGamma:1984eoj} & 29.0 & 8 & 0.82 \\
MARK II \cite{Schellman:1984yz} & 29.0 & 21 & 1.11 \\
TASSO \cite{TASSO:1984nda} & 34.0 & 14 & 0.65 \\
TASSO \cite{TASSO:1989jyt} & 34.5 & 14 & 0.83 \\
TASSO \cite{TASSO:1989jyt} & 35.0 & 14 & 2.08 \\
CELLO \cite{CELLO:1989adw} & 35.0 & 9 & 0.19 \\
TASSO \cite{TASSO:1989jyt} & 42.6 & 14 & 2.14 \\
TOPAZ \cite{TOPAZ:1994voc} & 58.0 & 4 & 0.44 \\
ALEPH \cite{ALEPH:1996oqp} & 91.2 & 16 & 0.47 \\
ALEPH \cite{ALEPH:1999udi} & 91.2 & 14 & 1.36 \\
DELPHI \cite{DELPHI:1994qgk} & 91.2 & 13 & 0.81 \\
OPAL \cite{OPAL:1999zfe} & 91.2 & 16 & 0.51 \\
SLD \cite{SLD:1998coh} & 91.2 & 9 & 0.44 \\
DELPHI \cite{DELPHI:2000ahn} & 183.0 & 3 & 17.05 \\
DELPHI \cite{DELPHI:2000ahn} & 189.0 & 4 & 4.80 \\
SLD(UDS) \cite{SLD:1998coh} & 91.2 & 9 & 1.56 \\
SLD(C) \cite{SLD:1998coh} & 91.2 & 9 & 1.25 \\
SLD(B) \cite{SLD:1998coh} & 91.2 & 9 & 1.95 \\
BESIII \cite{BESIII:2022zit} & 2.23 & 9 & 2.38 \\
BESIII \cite{BESIII:2022zit} & 2.4 & 10 & 1.95 \\
BESIII \cite{BESIII:2022zit} & 2.8 & 12 & 0.67 \\
BESIII \cite{BESIII:2022zit} & 3.05 & 13 & 1.01 \\
BESIII \cite{BESIII:2022zit} & 3.4 & 14 & 0.95 \\
BESIII \cite{BESIII:2022zit} & 3.67 & 15 & 1.90 \\
\midrule
TOTAL                             &                &  315 &  1.47\\ 
\bottomrule
\end{tabularx}
\label{tab:my_table2} 
\end{table}

\begin{table}[!h]
\centering
\caption{Same as Tab.\Ref{tab:my_table1}, but for $\eta$.}
\begin{tabularx}{0.45\textwidth}{@{}lCCCc@{}} 
\toprule
Exp($\eta$)                          & $\sqrt{s}$ [GeV]    & $N_{dp}$   & $\chi^2/N_{dp}$\\ 
\midrule
ARGUS \cite{ARGUS:1989orf}           &9.46            & 6    & 5.69\\
HRS \cite{HRS:1987aky}             &29.0            & 13   & 3.10\\
MARK II \cite{Wormser:1988ru}      &29.0            & 7    & 0.56\\
JADE \cite{JADE:1985bzp}           &34.4            & 2    & 3.77\\
JADE \cite{JADE:1989ewf}           &35.0            & 4    & 0.44\\
CELLO \cite{CELLO:1989byk}         &35.0            & 4    & 0.18\\
ALEPH \cite{ALEPH:1992zhm}         &91.2            & 8    & 0.59\\
ALEPH \cite{ALEPH:1999udi}         &91.2            & 18   & 1.07\\
ALEPH \cite{ALEPH:2001tfk}         &91.2            & 5    &11.18\\
L3 \cite{L3:1992pbe}               &91.2            & 5    & 1.11\\
L3 \cite{L3:1994gkb}               &91.2            & 11   & 1.19\\
OPAL \cite{OPAL:1998enc}           &91.2            & 11   & 0.90\\
BESIII \cite{BESIII:2024hcs}       &2.00            & 8    & 2.38\\
BESIII \cite{BESIII:2024hcs}       &2.20            & 9    & 0.67\\
BESIII \cite{BESIII:2024hcs}       &2.39            & 10   & 1.25\\
BESIII \cite{BESIII:2024hcs}       &2.64            & 11   & 0.36\\
BESIII \cite{BESIII:2024hcs}       &2.90            & 13   & 0.67\\
BESIII \cite{BESIII:2024hcs}       &3.05            & 13   & 0.55\\
BESIII \cite{BESIII:2024hcs}       &3.50            & 15   & 0.72\\
BESIII \cite{BESIII:2024hcs}       &3.67            & 15   & 1.42\\
\midrule
TOTAL                             &                &  188 &  1.52\\
\bottomrule
\end{tabularx}
\label{tab:my_table3} 
\end{table}

\subsection{Analysis of BESIII datasets alongside our best-fit results}
In order to illustrate the impact of the inclusion of higher twist effect, in Figs. \Ref{fig:pi0_history}, \Ref{fig:ks_history}, \Ref{fig:eta_history}, we show the low-energy behavior of our fit for $\pi^0$, $K_S^0$, and $\eta$ mesons and compare them to the most recent determinations available in the literature, namely the AKK08 \cite{Albino:2008fy} for $K_S^0$,  and the AESSS \cite{Aidala:2010bn} for $\eta$, and the $\pi^0$ FFs will be compared with NNFF \cite{Bertone:2017tyb} and MAPFF \cite{AbdulKhalek:2022laj}.

For $\pi^0$ mesons, various theoretical predictions extrapolated from different FFs determined from existing world data are depicted in Fig. \Ref{fig:pi0_history}. The FFs are obtained with slightly different assumptions and show the sensitivity of the predictions to assumptions about the behavior at low-$z$ and different $\sqrt{s}$. Both AKRS \cite{Anderle:2016czy} and NNFF \cite{Bertone:2017tyb} are derived from inclusive annihilation data at NNLO accuracy, with AKRS \cite{Anderle:2016czy} incorporating small-$z$ resummation and NNFF \cite{Bertone:2017tyb} applying hadron mass corrections. The NNLO analysis by MAPFF \cite{AbdulKhalek:2022laj} encompasses low-$Q^2$ data derived from the lepton-proton fixed-target experiments at HERMES and COMPASS. Each analysis adopts distinct initial evolution scales and kinematic criteria, with NNFF \cite{Bertone:2017tyb} and MAPFF \cite{AbdulKhalek:2022laj} setting $Q_0=5$ GeV, whereas AKRS \cite{Anderle:2016czy} utilizes $Q_0=10.54$ GeV. At varying center-of-mass energies, we observed a reasonable agreement between our NNLO+HT fit and experimental data points. Particularly at lower energies, our best-fit demonstrates significant congruence with BESIII data \cite{BESIII:2022zit}, highlighting the higher twist contribution in detailing the $\pi^0$ production mechanism. 
Notably, the fit in the low $z$ domain reveals substantial enhancements.

In the analysis of $K_S^0$ mesons, Fig. \ref{fig:ks_history} illustrates a comparison between the BESIII datasets \cite{BESIII:2022zit} and our best-fit results, alongside the predictions from AKK08 \cite{Albino:2008fy} at NLO accuracy. The AKK08 fragmentation functions are extracted from $K_S^0$ production in single inclusive annihilation  with $\sqrt{s}$ ranging from 14 to 189 GeV, and in proton-proton collisions at $\sqrt{s}$ of 200, and 630 GeV, whose energy scales exceed the typical c.m. energies at BESIII. The AKK08 FFs adopt a minimum value for $z$ at 0.05 and an initial scale of $\mu_0 = \sqrt{2}$ GeV. To incorporate hadron mass corrections to enhance the accuracy of their predictions, AKK08 FFs use a fitted parameter rather than the actual mass value. However, AKK08 predictions struggle to describe the BESIII data.

Fig.\Ref{fig:eta_history} displays our fitting results for $\eta$ mesons depicted by the red curve, while the green dotted curve represents the theoretical predictions using the $\eta$ FFs from the AESSS \cite{Aidala:2010bn} parametrization at NLO. The AESSS FFs, derived from data of $\eta$ production in $e^+e^-$ annihilation with $\sqrt{s}$ approximately at 10, 30, and 90 GeV, and in proton-proton collisions at $\sqrt{s}$ about 200 GeV, adopt a $z$ minimum cut at $z_{min}=0.1$. Meanwhile, our fit incorporates data points down to $z \geq 0.05$. Evidently, the AESSS FFs have poor description about the BESIII data sets, and the agreement between its theoretical prediction and data tends to worsen with decreasing c.m. energy.

\subsection{Analysis of world data at high energies alongside our best-fit results}

Fig. \ref{fig:my_label1} displays old world data for $\pi^{\pm}$ and $\pi^0$ across various center-of-mass energies. For $\pi^{\pm}$ data, $D_i^{\pi^+}+D_i^{\pi^-}=2D_i^{\pi^0}$ is adopted. Our optimal NNLO fragmentation functions provide a good description of both $\pi^{\pm}$ and $\pi^0$ data. In comparison to the $\chi^2$ analyses of the $\pi^{\pm}$ experiments conducted by NNFF \cite{Bertone:2017tyb}, MAPFF \cite{AbdulKhalek:2022laj}, and AKRS \cite{Anderle:2016czy}, our $\chi^2$ analysis of $\pi^{\pm}$ yields consistent results.  Deviations between theoretical predictions and data are observed at large $z$ values for TASSO at 14 GeV, while a good agreement is achieved with TASSO at 34.0 GeV and 34.6 GeV.

Fig. \ref{fig:my_label2} shows comparisons between our best-fit FFs' theoretical predictions and $K_S^0$ production world data. The deviation between the theory and the data can be seen for the large value of $z$ for DELPHI 183 and DELPHI 189. These ﬁndings are consistent with the $\chi^2$ values listed in Tab.\Ref{tab:my_table2}. Similarly, these phenomena can be observed from  AKK08 \cite{Albino:2008fy}, and SAK20 \cite{Soleymaninia:2020ahn}. Tagged data from the SLD measurements and our theory predictions from our best-ﬁt FFs including mass corrections and higher twist effect agree with each other well. 

Comparison between the $\eta$ production datasets in SIA process analyzed in this study from diﬀerent experiments in world data and the corresponding theoretical predictions are shown in Fig.\ref{fig:my_label3}.  Comparing with the $\chi^2$ from the fit by AESSS \cite{Aidala:2010bn}, we arrive at similar conclusions regarding the contributions of errors across different experimental groups.

Generally, there is an overall good agreement between the experimental data from all experiments and our best-fit NNLO theoretical predictions, consistent with the individual $\chi^2$ values reported in Tab \ref{tab:my_table1}, \ref{tab:my_table2}, and \ref{tab:my_table3}. Remarkably, our theoretical predictions exhibit significant agreement with experimental data across a wide range of scale values, spanning from 2 GeV to more than 91 GeV. Additionally, our further investigation reveals that our fragmentation functions exhibit excellent performance for high collision energy ($>9$ GeV), even without requiring modifications to account for higher twist effects. This underscores that our adjustment for higher twist effects provides robust support in the low-energy region, without undermining the prevailing dominance of QCD factorization in the large energy regime.

\begin{figure*}[htbp]
    \centering
    \begin{minipage}{0.49\textwidth}
        \includegraphics[width=\linewidth]{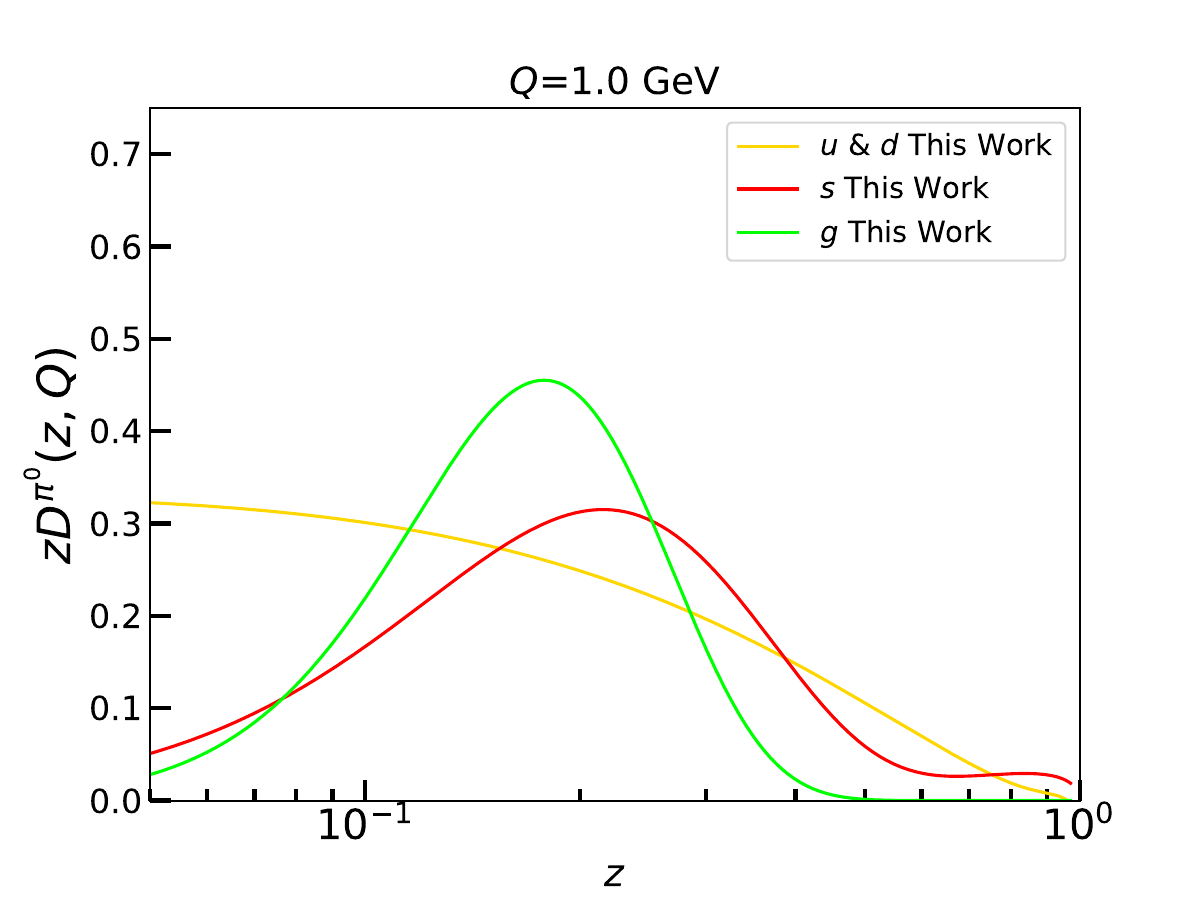}
    \end{minipage}\hfill
    \begin{minipage}{0.49\textwidth}
        \includegraphics[width=\linewidth]{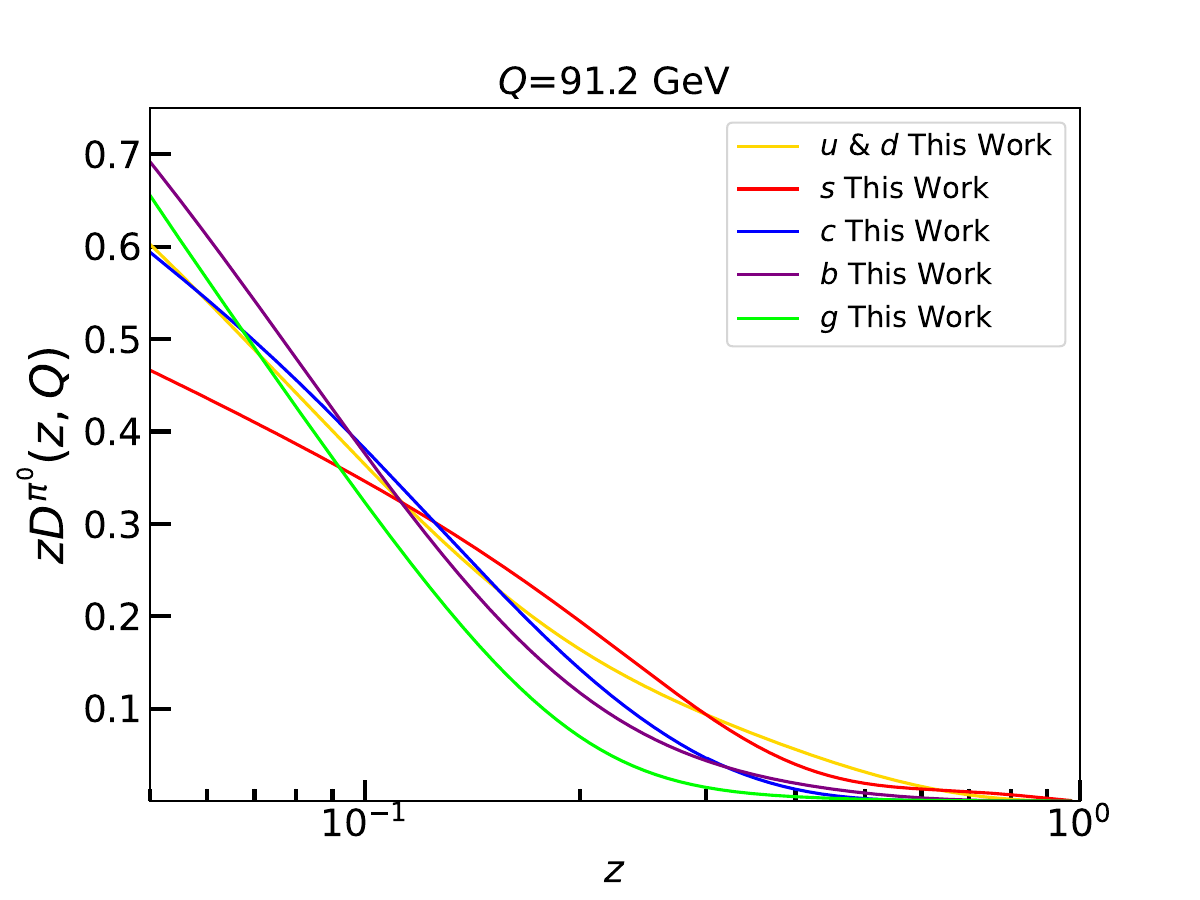}
    \end{minipage}
    \caption{Our best fit $\pi^0$ FFs at NNLO accuracy obtained for various partons at $\sqrt{s}=1.0$ GeV and $\sqrt{s}=91.2$ GeV.}
    \label{fig:FFs_pi0}
\end{figure*}

\begin{figure*}[htbp]
    \centering
    \begin{minipage}{0.49\textwidth}
        \includegraphics[width=\linewidth]{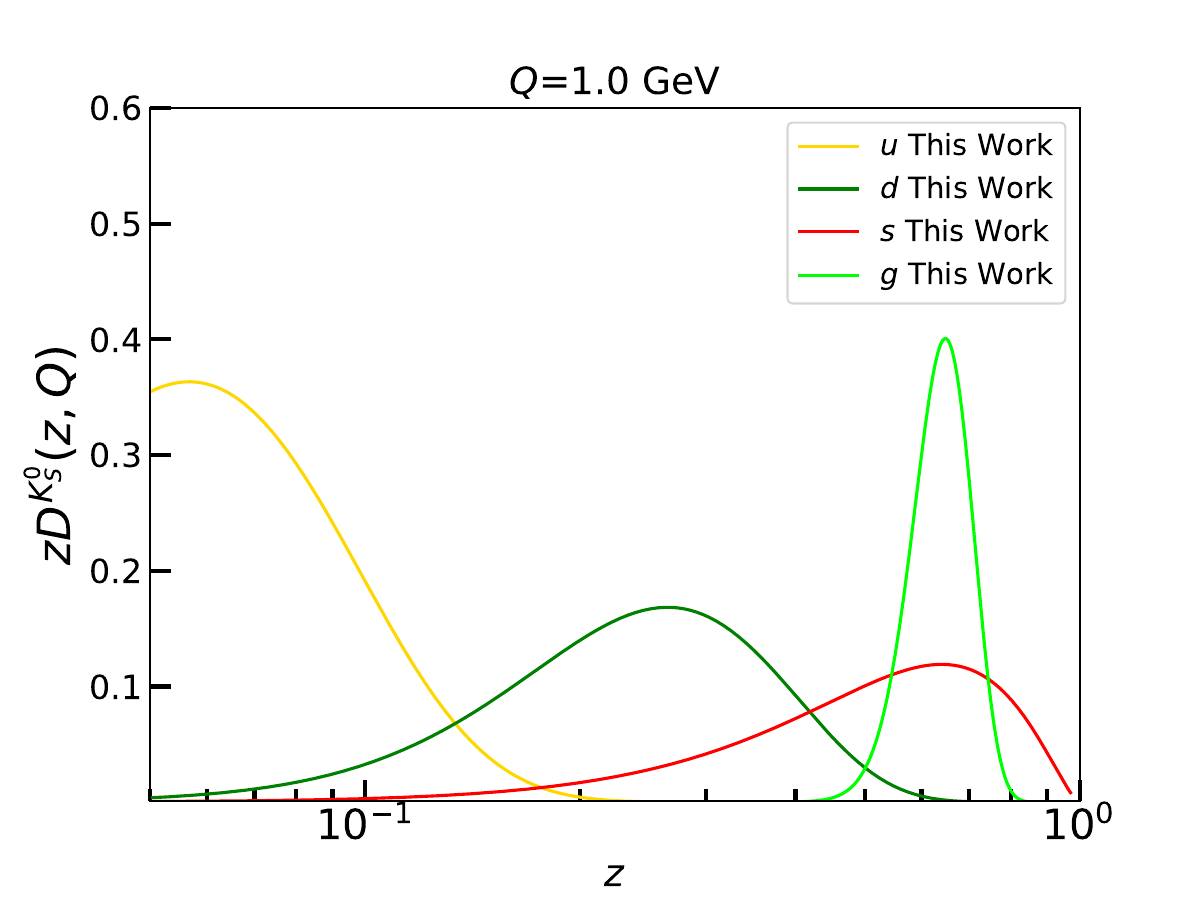}
    \end{minipage}\hfill
    \begin{minipage}{0.49\textwidth}
        \includegraphics[width=\linewidth]{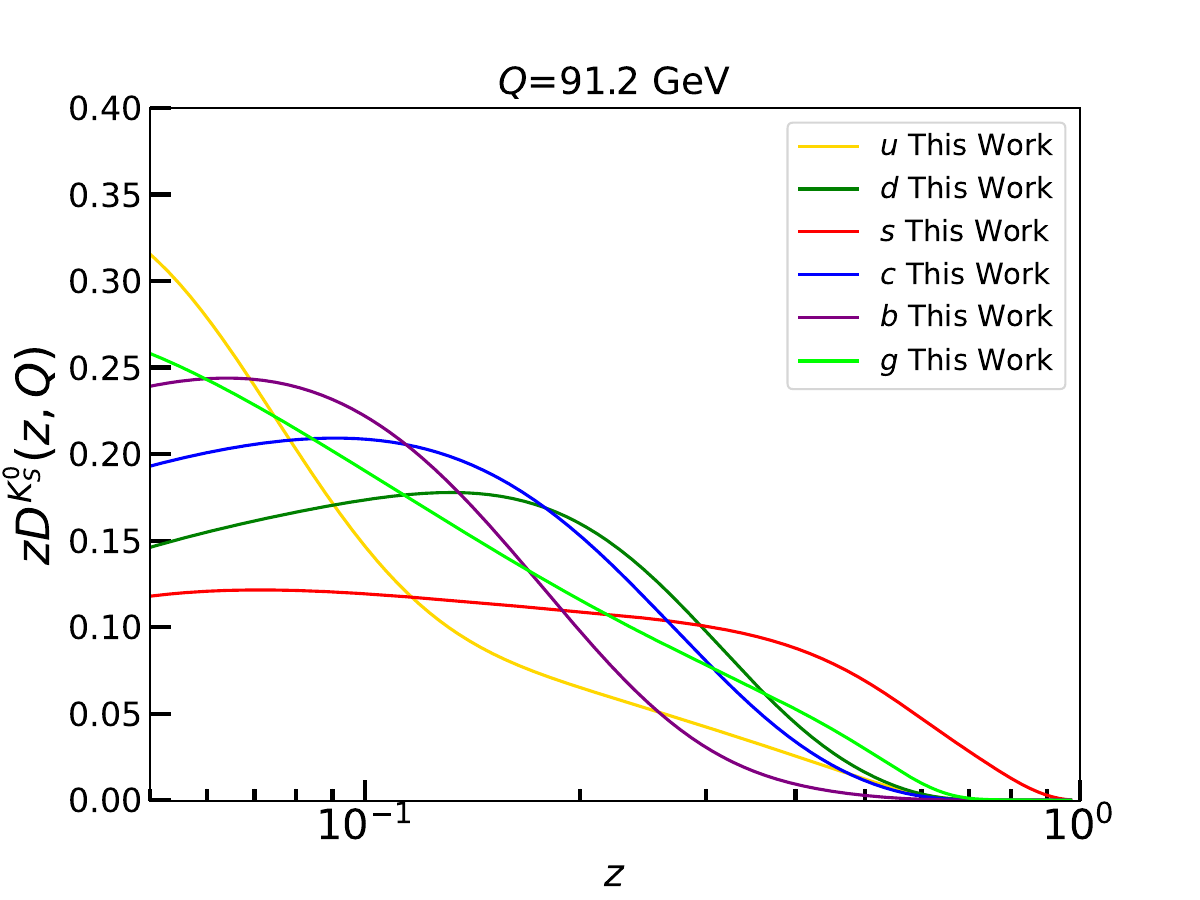}
    \end{minipage}
    \caption{Our best fit $K^0_S$ FFs at NNLO accuracy obtained for various partons at $\sqrt{s}=1.0$ GeV and $\sqrt{s}=91.2$ GeV.}
    \label{fig:FFs_ks}
\end{figure*}

\begin{figure*}[htbp]
    \centering
    \begin{minipage}{0.49\textwidth}
        \includegraphics[width=\linewidth]{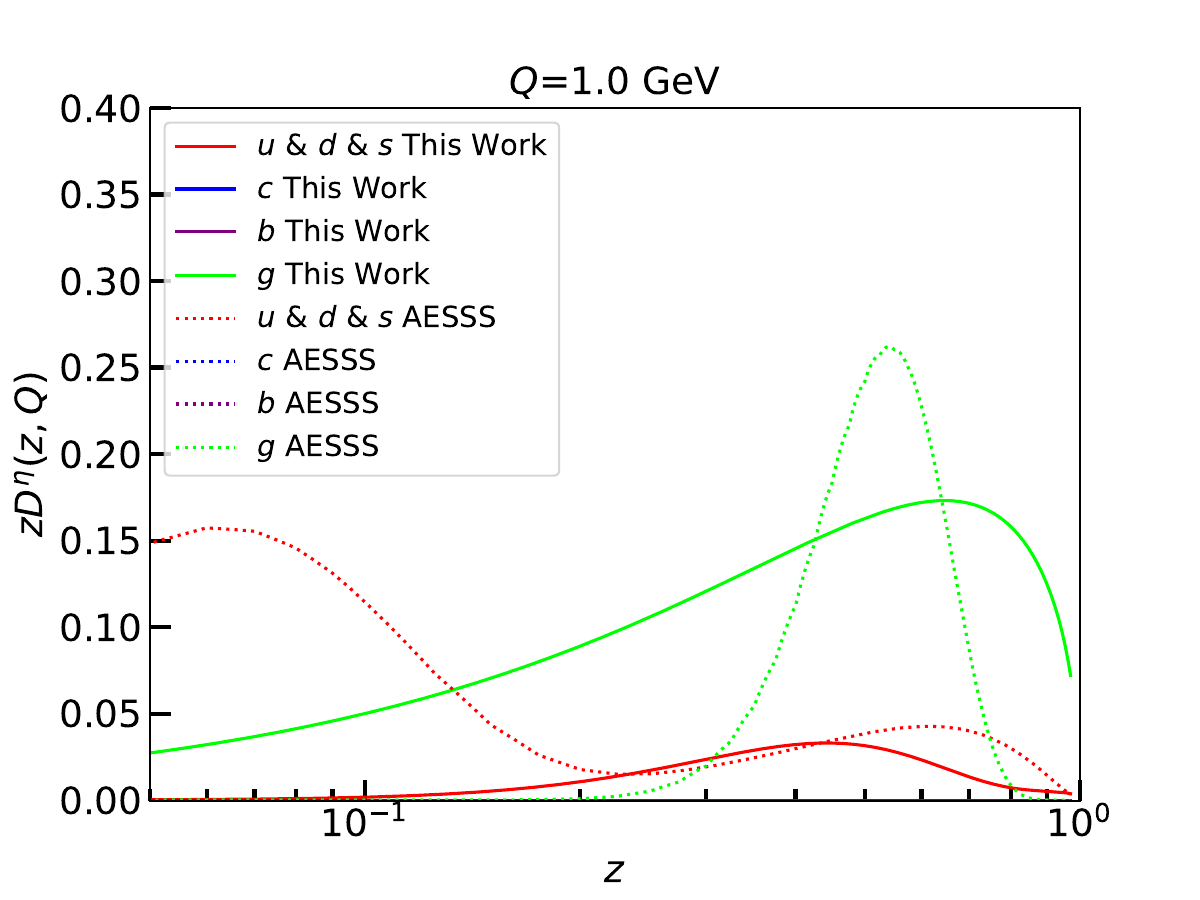}
    \end{minipage}\hfill
    \begin{minipage}{0.49\textwidth}
        \includegraphics[width=\linewidth]{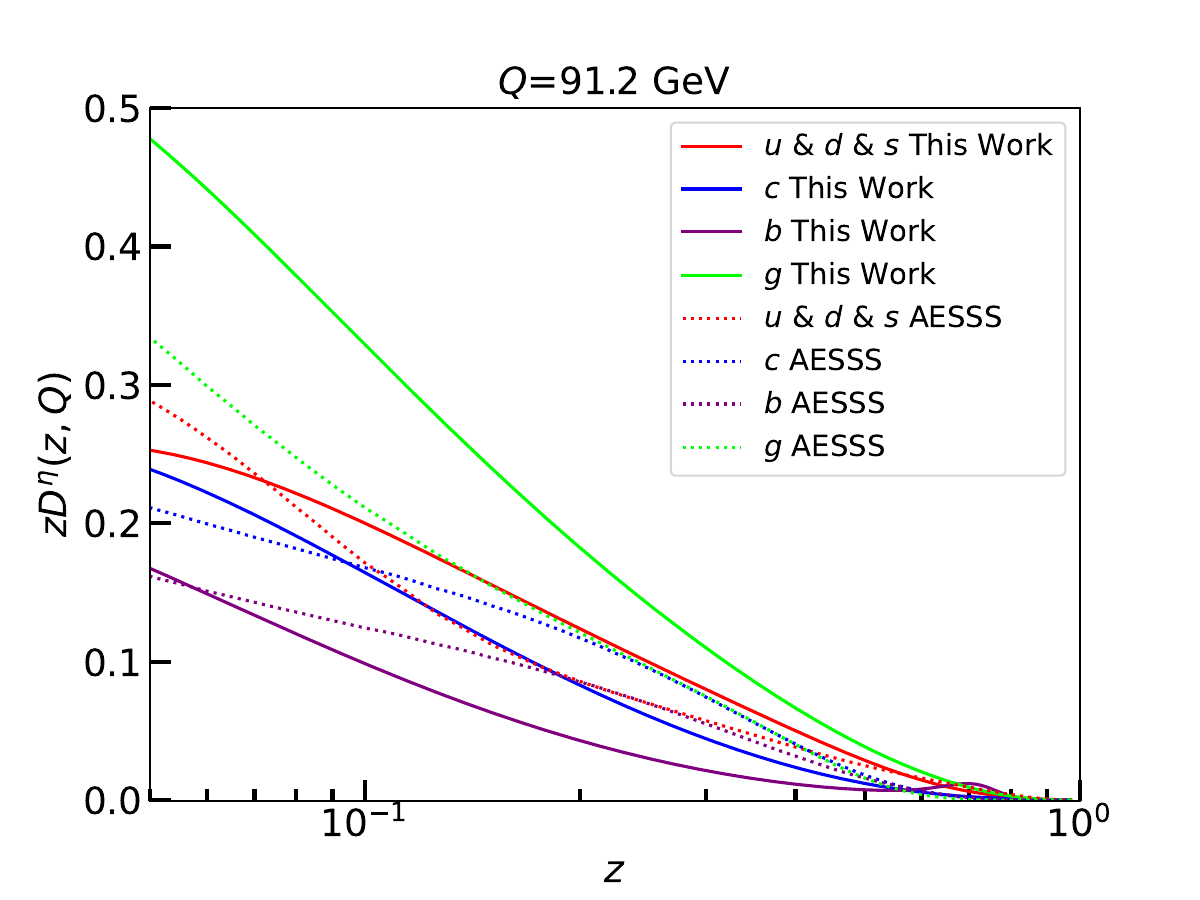}
    \end{minipage}
    \caption{Our best fit $\eta$ FFs at NNLO accuracy obtained for various partons at $\sqrt{s}=1.0$ GeV and $\sqrt{s}=91.2$ GeV. Results from AESSS \cite{Aidala:2010bn} FFs are also included for comparison.}
    \label{fig:FFs_eta}
\end{figure*}

The optimal fit parameters for $\pi^0$, $K^0_S$, and $\eta$ fragmentation functions (FFs), along with the corresponding higher twist effects, are detailed in Tables \ref{tab:parameter1}, \ref{tab:my_table5}, \ref{tab:my_table6}, and \ref{tab:eta_Q4}. Our obtained FFs are depicted in Figs. \Ref{fig:FFs_pi0}, \Ref{fig:FFs_ks}, \Ref{fig:FFs_eta}. 
To compare with the previously published FFs, we take AESSS \cite{Aidala:2010bn} as an example. Since it is the only available source for $\eta$ measurements, and given that it also employs an initial scale of $\mu_0=1$ GeV, we can directly compare our results with AESSS \cite{Aidala:2010bn} at both 1 GeV and 91.2 GeV in Fig. \ref{fig:FFs_eta}. At 1.0 GeV, up (u), down (d), and strange (s) quark FFs show similar trends between our fit and AESSS \cite{Aidala:2010bn} at high $z$ values, but diverge noticeably at low $z$ values. Significant differences are also observed in the gluon (g) FF at 1 GeV. At 91.2 GeV, the agreement between this work and AESSS \cite{Aidala:2010bn} improves across all quark types.
\begin{table*}[http]
\centering
\caption{Parameters  describing the NNLO FFs ($i^+=i+\bar{i}$) for $\pi^0$ with one standard deviation uncertainties. Inputs for light and gluon FFs are set at the initial scale $\mu_0 = 1.0$ GeV. Inputs for the charm and bottom FFs refer to $\mu=m_c$ and $\mu=m_b$, respectively.}
\label{tab:parameter1}
\begin{tabularx}{\textwidth}{@{}lCCCCC@{}}
\toprule
Parameter & N & $\alpha$ & $\beta$ & $\gamma$ & $\delta$ \\
\midrule
$u^+=d^+$ & $0.26 \pm 0.02$ & $-0.950 \pm 0.006$ & $1.92 \pm 0.05$ & $0.01 \pm 0.01$ & $180\pm 11$ \\
$s^+$ &$0.21 \pm 0.04$ & $1.40 \pm 0.04$ & $0.47 \pm 0.03$  & $1007 \pm 487$ & $8.4 \pm 0.7$ \\
$g$ & $0.07 \pm 0.01$ & $3.6 \pm 0.7$ & $19 \pm 12$ & $4000 \pm 2000$ & $2.4 \pm 1.7$ \\
$c^+$ & $0.21 \pm 0.02$ & $-0.09\pm 0.01$ & $8.2 \pm 0.5$ & $14.6 \pm 6.6$ & $106 \pm 6$ \\
$b^+$ & $0.22 \pm 0.01$ & $-0.7 \pm 0.1$ & $3.66 \pm 0.04$ & $10.1\pm 2.5$ & $8.5 \pm 1.0$ \\

\bottomrule
\end{tabularx}
\end{table*}
\begin{table*}[http]
\centering
\caption{Same as Tab.\ref{tab:parameter1}, but for $K_S^0$.}
\label{tab:my_table5}
\begin{tabularx}{\textwidth}{@{}lCCCCC@{}}
\toprule
Parameter & N & $\alpha$ & $\beta$ & $\gamma$ & $\delta$ \\
\midrule
$u^+$ & $0.058 \pm 0.005$ & $2.0 \pm 0.3$ & $50 \pm 34$ & - & - \\
$d^+$ & $0.10 \pm 0.04$ & $2.8 \pm 1.7 $ & $10.6 \pm 4.7$& - & - \\
$s^+$ & $0.12 \pm 0.03$ & $1.8 \pm 1.2$ & $1.6 \pm 0.4$ & - & - \\
$g$ & $0.061 \pm 0.005$ & $37.9 \pm 2.76$ & $21.0 \pm 2.1$ & - & - \\
$c^+$ & $0.101 \pm 0.007$ & $1.1 \pm 0.4$ & $8.8 \pm 1.0$ & - & - \\
$b^+$ & $0.078 \pm 0.007$ & $0.7 \pm 0.4$ & $14.2 \pm 3.0$ & - & - \\

\bottomrule
\end{tabularx}
\end{table*}
\begin{table*}[http]
\centering
\caption{Same as Tab.\ref{tab:parameter1}, but for $\eta$.}
\label{tab:my_table6}
\begin{tabularx}{\textwidth}{@{}l*{5}{C}@{}}
\toprule
Parameter & N & $\alpha$ & $\beta$ & $\gamma$ & $\delta$ \\
\midrule
$u^+=d^+=s^+$ & $0.030 \pm 0.003$ & $2.4 \pm 0.9$ & $0.4 \pm 0.7$ & $341 \pm 339$ & $4.0 \pm 0.5$ \\
$g$ & $0.123 \pm 0.005$  & $-0.1 \pm 0.4$ & $0.5 \pm 0.3$ & - & - \\
$c^+$ & $0.0008 \pm 0.0010$ & $46.8 \pm 41.7$   & $7.9 \pm 8.3$ & - & - \\
$b^+$ & $0.007 \pm 0.001$ & $50.0 \pm 43.1$   & $18.6 \pm 1.1$ & - & - \\
\bottomrule
\end{tabularx}
\end{table*}

\begin{table*}[htbp]
\caption{Parameters of higher twist effect for $\pi^0$, $K_S^0$, and $\eta$ with one standard deviation uncertainties.}
    \centering
    \small
    \begin{tabular}{|c|c|c|c|c|c|c|}
        \hline
        HT& $h_0$ & $h_1$ & $h_2$ & $h_3$ & $h_4$ & $h_5$ \\
        \hline
        $\pi^0$ & $-4.6 \pm 0.2$ & $-0.84 \pm 0.02$ & $-6.6 \pm 0.09$ & $6.2 \pm 0.4$ & $-1.46 \pm 0.04$ & $-8.0 \pm 0.1$ \\
        \hline
        $K_S^0$ & $-3.8 \pm 0.1$ & $-1.12 \pm 0.01$ & $-1.90 \pm 0.01$ & $1.5 \pm 0.2$ & $-2.60 \pm 0.05$ & $-2.28 \pm 0.09$ \\
        \hline
       $\eta$ & $-4.5 \pm 0.9$ & $-1.5 \pm 0.2$ & $-5.5 \pm 0.5$ & $3.0 \pm 0.8$ & $-2.8 \pm 0.5$ & $-6.8 \pm 1.7$\\
        \hline
    \end{tabular}

    \label{tab:eta_Q4}
\end{table*}

\section{SUMMARY AND OUTLOOK}
\label{sec-summary}
Our analysis has revealed challenges in simultaneously describing behaviors at low and high-energy scales using conventional fragmentation function approaches, underscoring the importance of testing the QCD factorization operational region. 
An exploratory study has been performed to understand it at NNLO level by including hadron mass corrections and higher-twist effects. 
To investigate the higher-twist contributions, we adopted a parameterized functional approach, creating an extensive framework that represents experimental outcomes across diverse energy scales, thus broadening classical theoretical models to encompass the BESIII regime.
This research establishes a framework for further exploring the interplay between higher twist dynamics and the hadronization process. Subsequent studies will focus on quantifying these effects and expanding the analysis to include a wider variety of hadronic states.
Despite SIA is the cleanest process for the determination of FFs, it carries limitted information on flavour separation, lacks the ability to distinguish between quark and antiquark FFs. To address these limitations, forthcoming updates to our fit will incorporate measurements from additional processes, such as proton-proton collisions and SIDIS processes, enhancing our grasp on these critical aspects.

\acknowledgments{
The authors are grateful for the useful discussions with Wenbiao Yan, Weiping Wang, Yateng Zhang, and Jian Zu.
This work is supported by the Strategic Priority Research Program of the Chinese Academy of Sciences under grant number XDB34000000, the Guangdong Major Project of Basic and Applied Basic Research Nos. 2020B0301030008 and 2022A1515010683.
HX is supported by the NSFC under Grants Nos. ~12035007 and 12022512.
YZ is supported by the NSFC under Grant No.~U2032105.
}


\end{document}